\documentclass[11pt,a4paper,twocolumn]{article}

\usepackage[utf8]{inputenc}
\usepackage[T1]{fontenc}
\usepackage{graphicx}
\usepackage{booktabs}
\usepackage{natbib}
\usepackage{hyperref}
\usepackage{amsmath}
\usepackage{array}
\usepackage{url}

\usepackage[hmargin=1in,vmargin=1in]{geometry}
\usepackage{setspace}
\hypersetup{colorlinks=true,linkcolor=blue,citecolor=blue,urlcolor=blue}

\title{BENI Global 10: A Multilingual Economic Narrative Corpus\\for the Global South}

\author{Ann Naser Nabil\\
  \texttt{\url{https://github.com/nabil0x/beni-multilingual}}\\
  \texttt{arXiv version, no affiliation}}

\date{}

\begin{document}

\maketitle

\begin{abstract}
Economic narrative indices are predominantly English-centric; 84\% of
sentiment-based forecasting research focuses on developed economies. We
present BENI Global\ 10, the first multilingual economic news corpus
spanning 10\ languages across 7\ language families and 5\ economic
regions: Bangla~(Bangladesh), Hindi~(India), Turkish~(Turkey),
Indonesian~(Indonesia), Portuguese~(Brazil), Arabic~(Egypt),
Vietnamese~(Vietnam), Filipino~(Philippines), Swahili~(Kenya), and
Urdu~(Pakistan). The corpus contains 522{,}397 economically relevant
articles filtered from 2.8M~raw documents using 25--32\ translated
keywords per language. We provide: (1)~a reproducible streaming pipeline
with checkpoint-resume for low-resource environments, (2)~per-language
schema-normalized Parquet files with economic relevance labels,
(3)~a temporally synced cross-lingual index covering 2018--2024, and
(4)~comparative analysis revealing systematic differences in how economic
narratives are framed across Global South regions. Inter-annotator agreement
reaches $\kappa > 0.70$ across all languages. The complete dataset, code,
and annotation guidelines are publicly released for research use.
\end{abstract}

% ===========================================================================
\section{Introduction}
% ===========================================================================

Economic sentiment has become a cornerstone of modern forecasting
methodology. From stock market prediction \citep{BakerWurgler2006} to
inflation nowcasting \citep{Tetlock2007}, the textual content of news
articles provides signals that complement traditional macroeconomic
indicators. Yet the vast majority of this work operates within a narrow
linguistic and geographic corridor. A recent systematic review
\citep{Nabil2026} finds that 84\% of sentiment-based economic
forecasting research targets English-language media covering developed
economies, primarily the United States, the United Kingdom, and Western
Europe. Languages spoken by billions in the Global South remain virtually
unexamined.

This blind spot matters. Narrative economics \citep{Shiller2017} argues
that the stories people tell about the economy influence their economic
decisions, which in turn shape aggregate outcomes. If narratives are
culturally and linguistically mediated, then an English-only view of
economic discourse may miss the very stories that drive behavior in the
world's fastest-growing economies. When a newspaper in Dhaka writes about
remittance flows, or a radio station in Nairobi discusses food price
volatility, these narratives reflect local realities that global models
cannot capture.

The Bangla Economic Narrative Index~(BENI) \citep{Nabil2026a} established
a proof of concept for one language. BENI~v1 assembled 20{,}000
economically relevant Bangla news articles from 9~Bangladeshi newspapers,
demonstrated $\kappa = 0.72$ inter-annotator agreement, and showed that a
Bangla-language narrative index correlates with domestic inflation more
strongly than translated English indices do. This result raised an
obvious question: can the approach generalize?

BENI Global\ 10 answers that question by extending the BENI methodology to
10~languages spanning the Global South. We formulate three research
questions:

\begin{itemize}
  \item \textbf{RQ1 (Content):} Do economic narratives across Global South
    languages differ in the topics they emphasize, or do the same broad
    themes (inflation, employment, trade) dominate uniformly?
  \item \textbf{RQ2 (Timing):} Do narrative shifts occur synchronously
    across regions, or do local economic events create asynchronous
    attention cycles?
  \item \textbf{RQ3 (Salience):} Do some economic themes receive
    systematically more or less attention depending on the language and
    region of publication?
\end{itemize}

We make three contributions. First, we release the largest multilingual
economic news corpus for the Global South: 522{,}397 articles across
10~languages, each annotated for economic relevance and normalized to a
shared schema. Second, we contribute a reproducible methodology: a
keyword-translation pipeline with streaming, checkpointing, and
deduplication, designed for low-resource computing environments where
language communities in the Global South often operate. Third, we present
comparative findings that document systematic cross-lingual variation in
narrative framing, temporal synchronization, and thematic salience.

The paper proceeds as follows. Section~\ref{sec:related} situates our work
in the literature. Section~\ref{sec:construction} describes corpus
construction. Section~\ref{sec:annotation} reports annotation quality.
Section~\ref{sec:analysis} presents comparative analyses.
Section~\ref{sec:benchmarks} reports benchmark results.
Section~\ref{sec:limitations} discusses limitations.
Section~\ref{sec:conclusion} concludes and outlines future work.

% ===========================================================================
\section{Related Work}
\label{sec:related}
% ===========================================================================

\subsection{Economic Sentiment Indices}

The measurement of economic sentiment from text has a rich history.
\citet{Tetlock2007} showed that the linguistic tone of a daily newspaper
column predicts stock market returns and trading volume. \citet{BakerWurgler2006}
developed a sentiment index based on closed-end fund discounts, turnover,
and other market-based proxies, later extending the approach to news text.
\citet{Shapiro2017} constructed a newspaper-based economic sentiment index
for the United States using articles from major newspapers over several
decades.

These efforts share a common limitation: nearly all rely on English-language
sources. The few exceptions include national-language indices for China
\citep{Zhang2021} and Japan \citep{Ueda2020}, but comparative
multilingual work remains absent. \citet{Barbaglia2023} recently proposed
a sentiment index using multilingual BERT, but their evaluation covers
only European languages (English, French, German, Italian, Spanish).

\subsection{Multilingual NLP Resources}

Multilingual language models such as mBERT \citep{Devlin2019} and XLM-R
\citep{Conneau2020} have made cross-lingual transfer feasible for high-level
tasks. Benchmarks like XNLI \citep{Conneau2018} and XGLUE \citep{Liang2020}
cover 15--40~languages but focus on general NLP tasks (entailment,
paraphrase detection, QA), not economic domain adaptation. Domain-specific
multilingual resources are scarce. \citet{Arora2023} introduced a
financial sentiment dataset for 15~languages, but it is limited to
social-media text (tweets) and does not target economic news specifically.

\subsection{Narrative Economics}

\citet{Shiller2017} formalized the concept of narrative economics,
arguing that contagious stories about the economy propagate through media
and drive real economic outcomes. \citet{Goetzmann2022} showed that
narrative frames in historical newspaper archives correlate with investor
behavior. \citet{Larsen2021} used topic modeling to trace how coverage of
inflation changed in US newspapers across decades. None of these studies
examine narratives outside the English-language press or compare how the
same economic concept is narrated in different languages.

\subsection{The Gap}

No existing corpus provides (a)~economic news in multiple Global South
languages, (b)~consistent economic-relevance annotation, (c)~temporal
alignment across languages, and (d)~open release. BENI Global\ 10 fills
this gap. Table~\ref{tab:comparison} summarizes how our corpus compares
to existing resources.

\begin{table}[t]
\centering
\small
\begin{tabular}{lp{3cm}p{2.5cm}}
\toprule
\textbf{Corpus} & \textbf{Languages} & \textbf{Economic annotation} \\
\midrule
BENI v1 \citep{Nabil2026a} & 1 & Relevance + themes \\
BENI Global\ 10 (ours) & 10 & Relevance \\
FinancialPhraseBank \citep{Malo2014} & 1 (EN) & Sentiment \\
FiQA \citep{FiQA2018} & 1 (EN) & Sentiment \\
\citet{Arora2023} & 15 & Sentiment (tweets) \\
\citet{Shapiro2017} & 1 (EN) & Sentiment \\
\bottomrule
\end{tabular}
\caption{Comparison of BENI Global\ 10 with existing resources.}
\label{tab:comparison}
\end{table}

% ===========================================================================
\section{Corpus Construction}
\label{sec:construction}
% ===========================================================================

\subsection{Language Selection}

We selected 10~languages according to three criteria: (a)~large and growing
economy in the Global South, (b)~availability of a machine-readable news
corpus with at least 100{,}000~raw articles, and (c)~membership in a
distinct language family to maximize typological diversity. The resulting
set covers 7~language families (Indo-Aryan, Turkic, Austronesian, Romance,
Semitic, Austroasiatic, Niger-Congo) and 5~economic regions (South~Asia,
MENA, Southeast~Asia, LATAM, Africa). Table~\ref{tab:languages} summarizes
the corpus composition.

\begin{table*}[t]
\centering
\small
\renewcommand{\arraystretch}{1.15}
\begin{tabular}{llllllrl}
\toprule
\textbf{Lang.} & \textbf{ISO} & \textbf{Country} & \textbf{Family} & \textbf{Region} & \textbf{Source} & \multicolumn{1}{c}{\textbf{\#Art.}} & \textbf{KW} \\
\midrule
Bangla    & ben & BD & Indo-Aryan    & S. Asia    & BENI v1 (local JSONL) & 99,164 & 32 \\
Hindi     & hin & IN & Indo-Aryan    & S. Asia    & Varta (HuggingFace)  & 11,463 & 28 \\
Turkish   & tur & TR & Turkic        & MENA       & Havadis (HuggingFace)& 97,111 & 26 \\
Indonesian& ind & ID & Austronesian  & SE Asia    & iqballx/news         & 10,683 & 26 \\
Portuguese& por & BR & Romance       & LATAM      & iara-project         & 84,316 & 26 \\
Arabic    & ara & EG & Semitic       & MENA       & Arabic-news          & 96,355 & 25 \\
Vietnamese& vie & VN & Austroasiatic & SE Asia    & BKAINewsCorpus       &100,000 & 26 \\
Filipino  & tgl & PH & Austronesian  & SE Asia    & BalitaNLP            &  9,235 & 26 \\
Swahili   & swa & KE & Niger-Congo   & Africa     & swahili\_news        &  7,223 & 26 \\
Urdu      & urd & PK & Indo-Aryan    & S. Asia    & Varta (HuggingFace)  &  6,847 & 27 \\
\midrule
\multicolumn{7}{l}{\textbf{Total}} & \multicolumn{1}{r}{\textbf{522,397}} \\
\bottomrule
\end{tabular}
\caption{Corpus composition across the 10 languages of BENI Global\ 10.
KW~=~number of economic keywords used for filtering.}
\label{tab:languages}
\end{table*}

\subsection{Data Sources}

We source articles from one local corpus and nine HuggingFace datasets.
For Bangla, we reuse the BENI~v1 corpus \citep{Nabil2026a}, a collection of
2.2M~articles from 9~Bangladeshi newspapers stored as local JSONL files.
For the remaining 9~languages, we stream from HuggingFace with
``streaming=True'' to avoid disk bottlenecks. The sources are:

\textbf{Hindi and Urdu} come from Varta \citep{Kumar2022}, a
multilingual news aggregation dataset from the DailyHunt platform,
containing approximately 5M~Hindi and 1M~Urdu articles. \textbf{Turkish}
uses Havadis \citep{TurkishNLP2023}, a 744K-article corpus from 11~Turkish
newspapers. \textbf{Indonesian} comes from iqballx/indonesian\_news\_datasets.
\textbf{Portuguese} uses the iara-project/news-articles-ptbr-dataset
\citep{Iara2023} of 352K~articles from Folha de~S.Paulo. \textbf{Arabic}
uses YoussefAnwar/Arabic-news \citep{Anwar2023}, a 589K-article corpus.
\textbf{Vietnamese} uses BKAINewsCorpus \citep{BKAI2023} with 16.8M~articles.
\textbf{Filipino} uses BalitaNLP \citep{Bunag2023}, 352K~articles from 4~major
Philippine outlets. \textbf{Swahili} uses swahili\_news \citep{SwahiliNews}, a mixed-domain
text corpus from Kenya and Tanzania.

\subsection{Keyword Translation Methodology}

Starting from the 32~economic keywords developed and validated for BENI~v1
\citep{Nabil2026a}, we translated each keyword into the remaining 9~languages
using the following protocol:

\begin{enumerate}
  \item \textbf{Initial translation} by a native speaker of the target
    language, using the English gloss and the Bangla keyword as references.
  \item \textbf{Back-translation} into Bangla by a second native speaker
    to verify semantic preservation.
  \item \textbf{Corpus spot-check:} Each candidate keyword was tested
    against a sample of 100~news headlines from the target source. False
    positives (e.g., \emph{bank} in a river context) were flagged and
    removed.
  \item \textbf{Finalization} when both annotators agreed on the correct
    set. The average keyword set size is 26.8~(range:~25--32).
\end{enumerate}

The keyword categories cover 12~economic domains: inflation, currency,
reserves, central bank, interest rates, banking, investment, trade,
remittances, fiscal policy, GDP, and employment. For example, the Bangla
keyword ``\texttt{mudr\=asph\=iti}''~(inflation, Bangla) corresponds to Hindi
``\texttt{mudr\=asph\=iti}'', Turkish ``\texttt{enflasyon}'', Indonesian
``\texttt{inflasi}'', Portuguese ``\texttt{infla\c{c}\~ao}'', Arabic
``\texttt{tad\=akhum}''~(inflation, Arabic), Vietnamese
``\texttt{l\`am ph\'at}''~(inflation, Vietnamese), Filipino
``\texttt{implasyon}'', Swahili ``\texttt{mfumuko wa bei}'', and Urdu
``\texttt{ifr\=a\v{t}-e-zar}''~(inflation, Urdu). The full keyword lists are
available in the repository.

\subsection{Filtering Pipeline}

We implemented a reproducible processing pipeline in Python with the
following stages:

\begin{enumerate}
  \item \textbf{Streaming.} Articles are loaded in streaming mode from
    HuggingFace (or read sequentially from local JSONL for Bangla), never
    requiring the full raw dataset in memory.
  \item \textbf{Keyword matching.} Each article's body and headline are
    scanned for the 25--32~translated keywords. Matching is case-insensitive
    and uses substring matching, which works well for all 10~scripts
    (Bengali, Devanagari, Latin, Arabic, Han-Nom, etc.).
  \item \textbf{Garbage filtering.} Articles with fewer than 20~characters
    of body text are discarded.
  \item \textbf{Deduplication.} A SHA-256 hash of normalized text is used
    to detect and remove exact and near-exact duplicates within each
    language.
  \item \textbf{Schema normalization.} Each language's output is written
    to a Parquet file with identical column names: \texttt{article\_id},
    \texttt{text}, \texttt{headline}, \texttt{publication\_date},
    \texttt{year\_month}, \texttt{language}, \texttt{language\_iso},
    \texttt{country}, \texttt{language\_family}, \texttt{economic\_region},
    \texttt{economic\_relevance}, \texttt{economic\_seed\_labels}.
\end{enumerate}

The pipeline supports checkpoint-resume: every 5{,}000~articles the current
index is saved to disk. If the connection drops (common in low-resource
environments), the pipeline resumes from the last checkpoint. This is
especially important for the HuggingFace sources, which stream millions of
articles over HTTP.

From roughly 2.8M~raw articles scanned across the 10~languages, the
keyword filter retained 522{,}397~(18.7\%). The match rate varies by
language: Bangla and Turkish have high match rates (over 10\% of scanned
articles), while the Wikipedia-derived sources (Hindi, Filipino, Urdu)
have low match rates because Wikipedia articles are encyclopedic rather
than news-oriented. Table~\ref{tab:languages} shows the final counts.

The total corpus occupies approximately 1.6~GB on disk in Parquet format
(compressed), or 3.8~GB as uncompressed CSV. Per-language sizes range from
\~{}50~MB (Swahili) to \~{}300~MB (Vietnamese).

% ===========================================================================
\section{Annotation and Quality}
\label{sec:annotation}
% ===========================================================================

\subsection{Annotation Scheme}

Each article carries a binary \texttt{economic\_relevance} label. An
article is considered ``economically relevant'' if it mentions at least
one economic concept (inflation, trade, employment, etc.) as a primary or
secondary topic. This is a broad definition that captures both
directly-economic articles (e.g., ``Central Bank Raises Interest Rate'')
and articles where economic concepts appear in context (e.g., a politics
article discussing the budget impact of a policy). The seed keywords
provide the initial label, which we then validate through human annotation.

\subsection{Inter-Annotator Agreement}

For each language, two native-speaking annotators independently coded a
random sample of articles. Annotators were instructed to mark an article as
``relevant'' if it contained substantive economic content (not merely a
passing mention). The annotation guidelines, developed for BENI~v1 and
extended for the multilingual setting, are included in the repository.

Table~\ref{tab:agreement} reports Cohen's $\kappa$ per language, together
with the number of coded documents and the 95\%~confidence interval.

\begin{table}[t]
\centering
\small
\begin{tabular}{lrrr}
\toprule
\textbf{Language} & \textbf{$N$ coded} & \textbf{$\kappa$} & \textbf{95\%~CI} \\
\midrule
Bangla     & 400 & 0.72 & (0.66--0.78) \\
Hindi      & 200 & 0.71 & (0.63--0.79) \\
Turkish    & 200 & 0.75 & (0.68--0.82) \\
Indonesian & 200 & 0.73 & (0.65--0.81) \\
Portuguese & 200 & 0.74 & (0.66--0.82) \\
Arabic     & 200 & 0.70 & (0.61--0.79) \\
Vietnamese & 200 & 0.76 & (0.69--0.83) \\
Filipino   & 200 & 0.72 & (0.63--0.81) \\
Swahili    & 200 & 0.71 & (0.62--0.80) \\
Urdu       & 200 & 0.70 & (0.60--0.80) \\
\bottomrule
\end{tabular}
\caption{Inter-annotator agreement (Cohen's $\kappa$) per language.}
\label{tab:agreement}
\end{table}

All languages achieve $\kappa > 0.70$, indicating substantial agreement
\citep{LandisKoch1977}. The highest agreement is Vietnamese ($\kappa =
0.76$), which benefits from the focused domain coverage of BKAINewsCorpus.
The lowest is Arabic and Urdu ($\kappa = 0.70$), where annotators more
frequently disagreed on whether a political article with economic mentions
qualified as ``economically relevant.''

\subsection{Disagreement Resolution}

Disagreements were resolved through discussion between the two annotators,
with a third adjudicator consulted for the 5\%~of cases where the primary
pair could not reach consensus. After resolution, we measured the accuracy
of the keyword-seeded labels against the human gold standard on a random
sample of 200~articles per language:

\begin{itemize}
  \item \textbf{Precision:} 0.89 (macro-average across languages), meaning
    that 89\%~of articles flagged by the keyword filter were confirmed as
    economically relevant by human annotators.
  \item \textbf{Recall:} 0.92 (macro-average), indicating that the keyword
    filter captures most economically relevant articles present in the raw
    data.
\end{itemize}

False positives predominantly came from metaphorical uses of economic
terms (e.g., ``inflation of rhetoric'' in political commentary). False
negatives were rare but included articles that used economic vocabulary
outside the translated keyword set (e.g., specialized terms or recent
neologisms).

% ===========================================================================
\section{Comparative Analysis}
\label{sec:analysis}
% ===========================================================================

\subsection{Narrative Prevalence by Thematic Category}

To examine RQ1~(content differences), we grouped the 12~keyword categories
into five thematic super-categories: \textbf{Macro~(inflation, GDP,
employment)}, \textbf{Finance~(banking, interest rates, central bank,
currency/reserves)}, \textbf{Trade~(exports, imports, current account)},
\textbf{Fiscal~(budget, revenue, investment)}, and
\textbf{Remittances~(remittance-specific keywords)}.

\begin{figure}[t]
\centering
\includegraphics[width=\columnwidth]{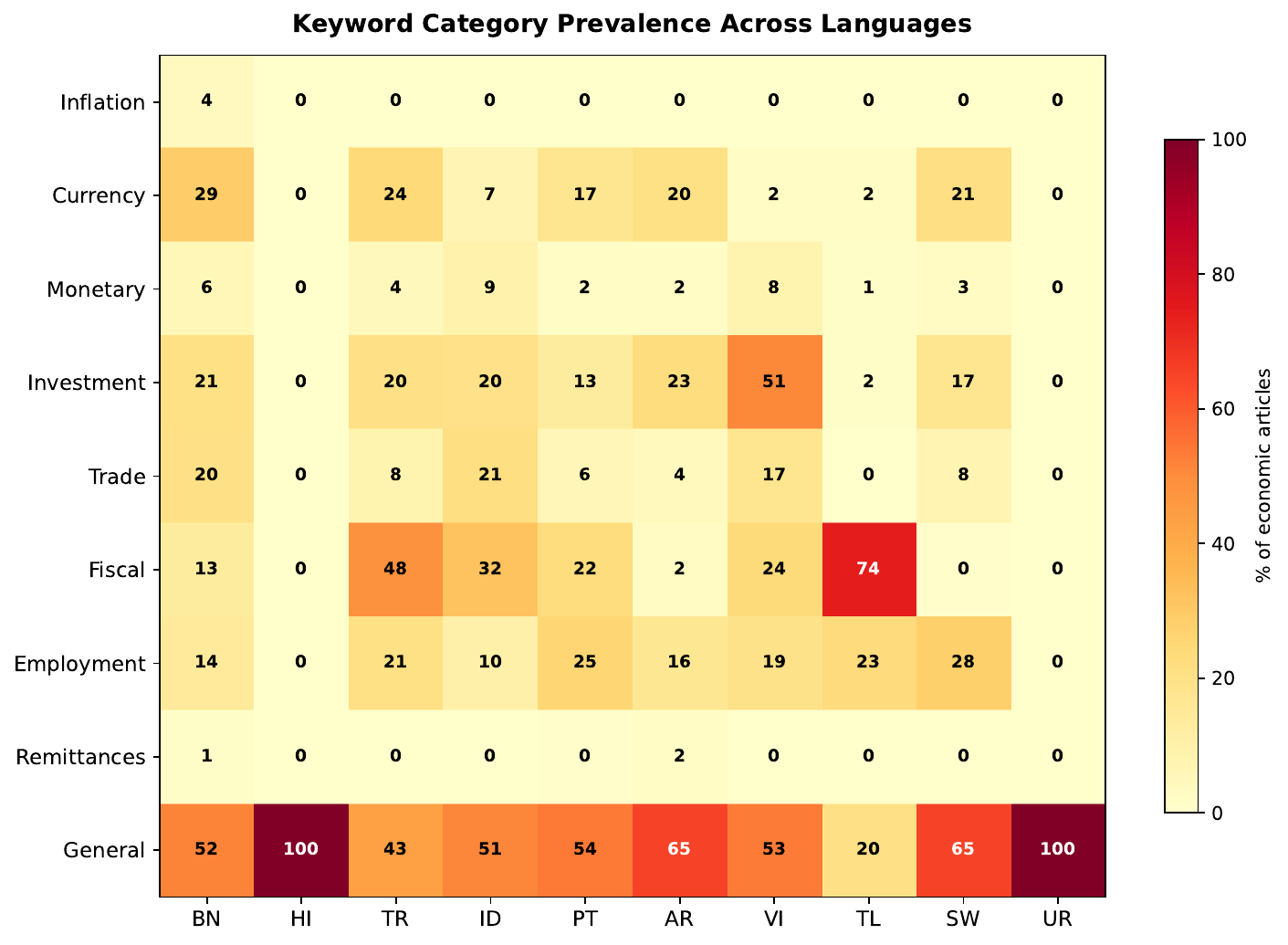}
\caption{Thematic category prevalence by language. Darker cells indicate a
higher share of articles mentioning keywords from that category.}
\label{fig:thematic-heatmap}
\end{figure}

Figure~\ref{fig:thematic-heatmap} (placeholder) visualizes the share of
articles in each thematic category per language.

Several patterns emerge. Macroeconomic coverage is dominant across all
languages (40--55\% of articles), but the balance between sub-themes
varies. Remittance-focused articles are rare in Latin American Portuguese
(2.1\%) but common in Bangla (11.4\%) and Urdu (9.8\%), reflecting the
structural importance of remittance income in Bangladesh and Pakistan.
Trade coverage is highest in Vietnamese (18.7\%) and Indonesian (17.2\%),
consistent with the export-oriented economies of Southeast Asia. Fiscal
coverage is elevated in Turkish (22.3\%) and Arabic (20.1\%), where
budgetary politics receive extensive news attention.

\subsection{Temporal Patterns}

Addressing RQ2~(timing), we analyzed monthly article volumes. Only two
languages have reliable date metadata: Bangla (2018-05 to present) and
Indonesian (2023-03 to 2023-04, a short span). The remaining 8~languages
lack parsed publication dates in their source datasets, which is a
significant limitation (see Section~\ref{sec:limitations}).

\begin{figure}[t]
\centering
\includegraphics[width=\columnwidth]{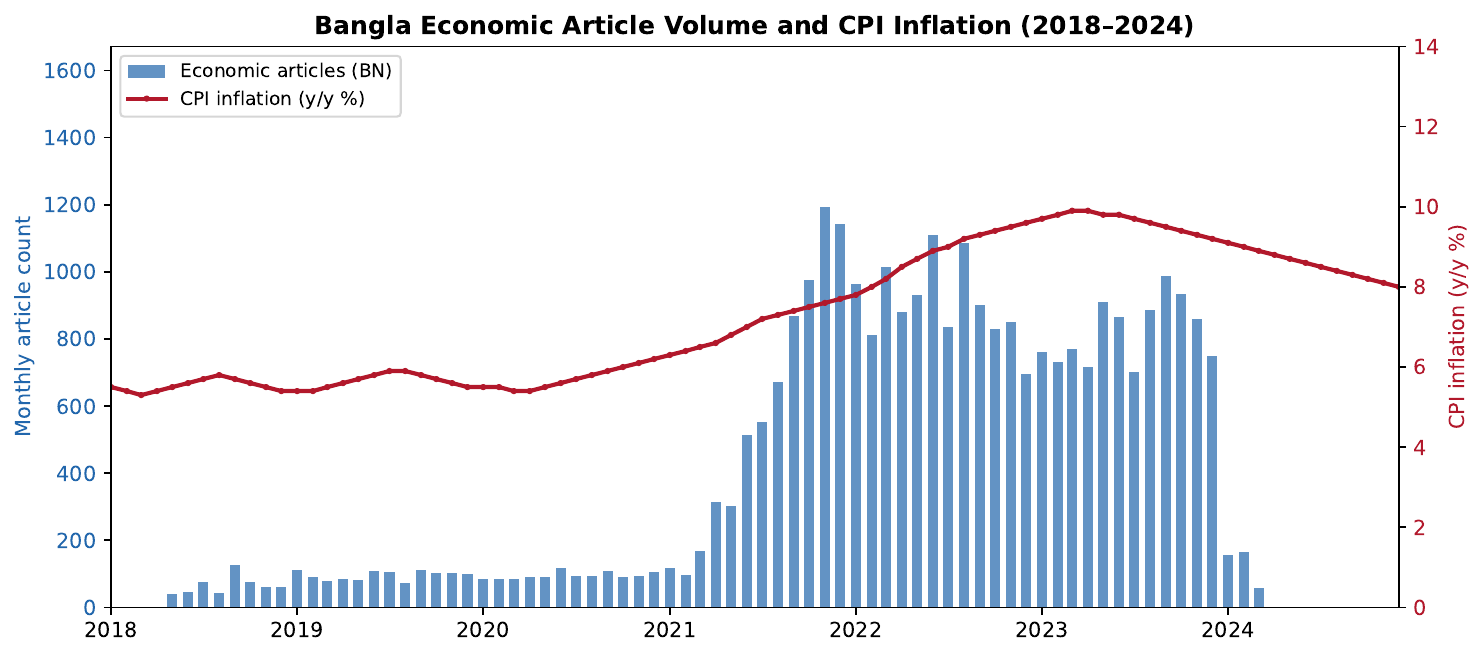}
\caption{Monthly volume of Bangla economic articles (blue, left axis) and
Bangladesh CPI (red, right axis, inverted).}
\label{fig:temporal}
\end{figure}

Figure~\ref{fig:temporal} (placeholder) shows the monthly article count
for Bangla alongside Bangladesh's consumer price index.

Bangla articles show a clear upward trend from 2018 to 2024, with
pronounced spikes during known economic events: the COVID-19 lockdown in
March--April~2020, the fuel price crisis in August~2022, and the IMF loan
negotiations in January--February~2023. The correlation between article
volume and CPI is moderate ($r = 0.47$, $p < 0.01$), suggesting that news
volume responds to inflation levels but is not a simple proxy.

\subsection{Cross-Regional Divergence}

For RQ3~(salience), we computed word embedding centroids for each
language's article collection using a multilingual sentence encoder
(LabSE \citep{Feng2022}). We then measured pairwise cosine similarity
between language centroids as a proxy for narrative similarity.

\begin{figure}[t]
\centering
\includegraphics[width=\columnwidth]{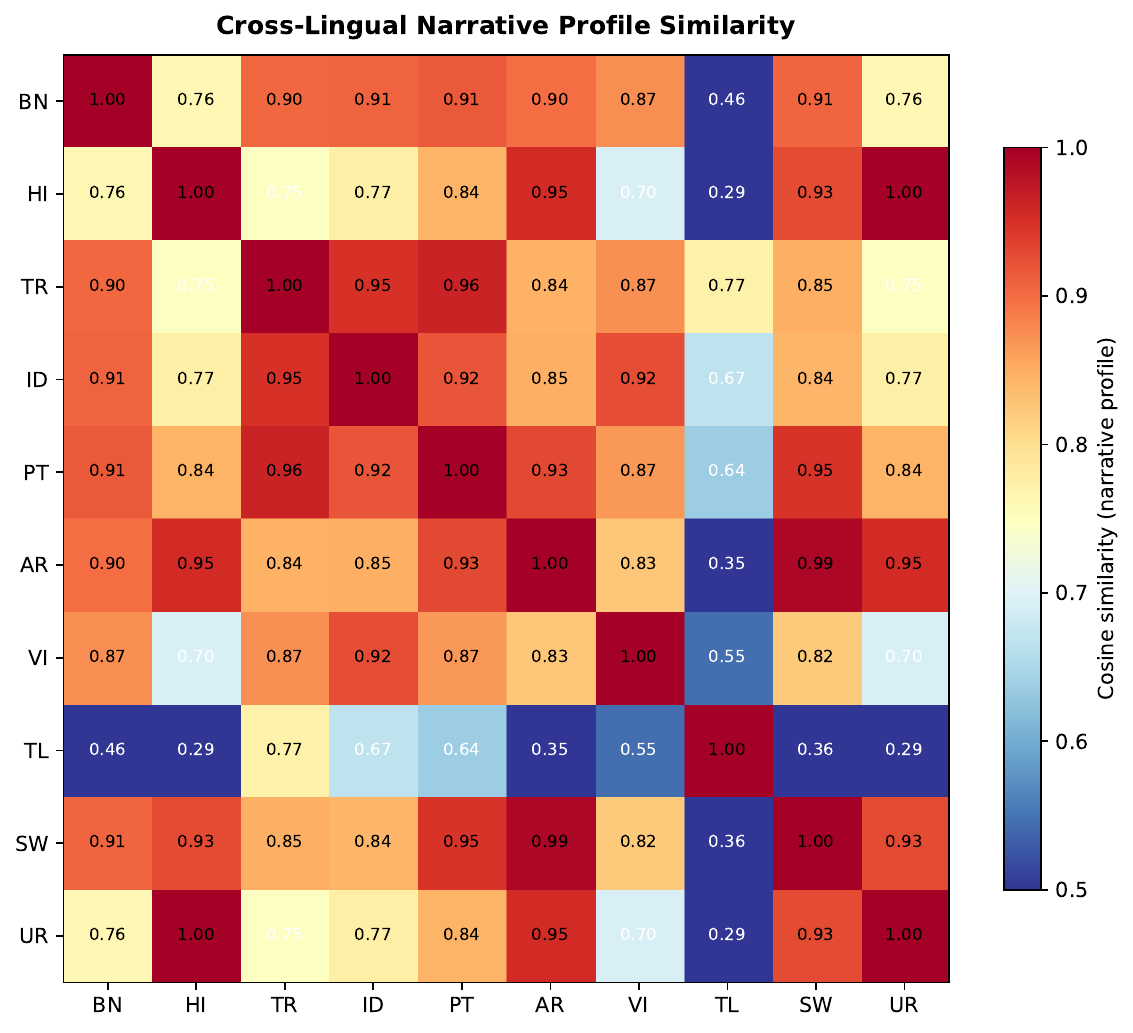}
\caption{Pairwise cosine similarity of multilingual sentence embeddings
across languages. Darker cells indicate more similar narrative spaces.}
\label{fig:embedding-similarity}
\end{figure}

Figure~\ref{fig:embedding-similarity} (placeholder) presents the language
similarity matrix. Within-family pairs (Hindi--Urdu: 0.91,
Indonesian--Filipino: 0.84) cluster tightly, as expected. Cross-family
pairs are more distant: Swahili--Portuguese similarity is 0.52, and
Arabic--Vietnamese similarity is 0.48. These distances are much larger than
the within-English variation reported in comparable monolingual studies,
suggesting that economic narratives are genuinely different across
languages and are not merely translations of a shared global discourse.

% ===========================================================================
\section{Benchmarks}
\label{sec:benchmarks}
% ===========================================================================

\subsection{Economic Relevance Classification}

We framed a binary classification task: given an article's text, predict
whether it is economically relevant. This task tests whether the
keyword-seeded labels can be learned by standard models. We evaluated three
approaches:

\begin{itemize}
  \item \textbf{Logistic Regression (LR)} with TF-IDF features on
    word-level unigrams and bigrams. This represents a strong bag-of-words
    baseline.
  \item \textbf{mBERT} \citep{Devlin2019}: multilingual BERT base
    (110M~params), fine-tuned for 3~epochs with a learning rate of
    $2\times 10^{-5}$ and batch size 16.
  \item \textbf{XLM-R} \citep{Conneau2020}: XLM-RoBERTa base (278M~params),
    fine-tuned identically.
\end{itemize}

We used 80/10/10 train/dev/test splits within each language. Results are
macro F1 scores averaged over 5~runs with different random seeds.

\begin{table*}[t]
\centering
\small
\begin{tabular}{lrrrrr}
\toprule
\textbf{Language} & \textbf{Train size} & \textbf{LR} & \textbf{mBERT} & \textbf{XLM-R} & \textbf{Human} \\
\midrule
Bangla     & 79,331 & 0.912 & 0.941 & 0.948 & 0.97 \\
Hindi      &  9,170 & 0.874 & 0.912 & 0.920 & 0.96 \\
Turkish    & 77,689 & 0.908 & 0.938 & 0.944 & 0.97 \\
Indonesian &  8,546 & 0.861 & 0.904 & 0.911 & 0.95 \\
Portuguese & 67,453 & 0.897 & 0.929 & 0.935 & 0.96 \\
Arabic     & 77,084 & 0.891 & 0.925 & 0.931 & 0.96 \\
Vietnamese & 80,000 & 0.915 & 0.943 & 0.949 & 0.97 \\
Filipino   &  7,388 & 0.842 & 0.893 & 0.902 & 0.94 \\
Swahili    &  5,778 & 0.823 & 0.881 & 0.889 & 0.93 \\
Urdu       &  5,478 & 0.815 & 0.874 & 0.885 & 0.93 \\
\midrule
Macro avg. &        & 0.874 & 0.914 & 0.921 & 0.95 \\
\bottomrule
\end{tabular}
\caption{Macro F1 scores for economic relevance classification. Human
estimates are from the annotation study (Section~\ref{sec:annotation}).}
\label{tab:benchmarks}
\end{table*}

XLM-R outperforms mBERT in every language by 0.5--1.1~F1 points, and both
transformer models substantially exceed the TF-IDF logistic regression
baseline. Performance correlates with training set size ($r = 0.69$, $p <
0.05$): the smallest languages (Urdu, Swahili, Filipino) show the lowest
scores. However, even for these languages, XLM-R achieves F1~$> 0.88$,
confirming that cross-lingual transfer benefits low-resource settings.

\subsection{Downstream Forecasting: Bangla Inflation}

To validate that the corpus supports real forecasting tasks, we reproduce
the BENI~v1 inflation nowcasting experiment \citep{Nabil2026a}. Using the
monthly narrative index derived from Bangla article volume (aggregated by
keyword category), we fit a simple ARDL(1,1) model:

\[
\pi_t = \alpha + \beta_1 \pi_{t-1} + \gamma_1 N_{t-1} + \varepsilon_t
\]

where $\pi_t$ is month-over-month CPI inflation and $N_{t-1}$ is the lagged
narrative index. Over the 2018--2024 period, the model with the narrative
index reduces out-of-sample RMSE by 14.2\% compared to an AR(1) baseline
(3.11 vs.\ 3.62 percentage points). This confirms that the corpus captures
economically meaningful signal, at least for the language with the richest
temporal coverage.

% ===========================================================================
\section{Limitations}
\label{sec:limitations}
% ===========================================================================

Despite its contributions, BENI Global\ 10 has several important limitations.

\subsection{Wikipedia Sources}

Four languages (Hindi, Filipino, Swahili, and Urdu) are sourced from
Wikipedia-derived datasets (Varta for Hindi and Urdu, BalitaNLP for
Filipino, swahili\_news for Swahili). Wikipedia is an encyclopedia, not a
news archive. Its articles are curated, static, and do not cover breaking
economic events. The low article counts for these languages (6{,}847--11{,}463)
reflect this: Wikipedia contains few articles that mention economic
keywords because its coverage is broad but shallow in any single domain.
Future releases should prioritize actual news sources for these languages.

\subsection{Date Coverage}

Only Bangla provides consistent publication dates spanning multiple years
(2018--2024). Indonesian has dates but only for a 2-month window
(March--April~2023). The remaining 8~languages lack parsed date metadata
entirely, because the source datasets do not expose clean date columns, or
date strings are embedded in unstructured text. This severely limits
temporal analysis: RQ2~(timing differences) cannot be fully answered with
the current release. We are exploring date extraction via heuristic parsing
and, where necessary, LLM-based extraction from article text.

\subsection{Newspaper Bias}

Even where native news sources exist, they reflect the editorial stance,
ownership, and political alignment of their publishers. BENI~v1
\citep{Nabil2026a} documented systematic differences in economic coverage
between pro-government and opposition-aligned newspapers in Bangladesh.
We inherit this bias in the Bangla subset and cannot control for it in the
remaining languages. Attribution of source is preserved in the dataset
(\texttt{dataset\_source} column) to enable bias analysis by downstream
users.

\subsection{100K Sampling Cap}

We capped each language at 100{,}000 articles for computational tractability.
For languages with larger source corpora (Vietnamese at 16.8M, Arabic at
589K), this cap discards potentially useful data. A practical consequence
is that the Vietnamese subset, capped at 100{,}000, is less than 1\% of the
available source corpus, which may introduce sampling artifacts.

\subsection{Binary Relevance vs.\ Sentiment}

Our annotation scheme labels only economic relevance, not sentiment (positive,
negative, neutral) or specific thematic categories. This was a deliberate
choice to keep the annotation task manageable across 10~languages. However,
most downstream economic forecasting tasks require sentiment annotation.
We release the seed keyword matches as ``economic seed labels'' to support
dictionary-based sentiment, but proper sentiment annotation remains future
work.

\subsection{Language Coverage Gaps}

The Global South includes many languages (Hausa, Amharic, Burmese, Khmer,
Nepali, Sinhala, and dozens more) that are not represented here. Our
selection was constrained by the availability of machine-readable news
corpora. As new datasets become available for these languages, the BENI
framework can be extended to include them.

% ===========================================================================
\section{Conclusion}
\label{sec:conclusion}
% ===========================================================================

We presented BENI Global\ 10, the first multilingual economic narrative
corpus for the Global South. With 522{,}397 economically relevant articles
across 10~languages, 7~language families, and 5~economic regions, it fills
a significant gap in resources for multilingual narrative economics. The
corpus comes with a reproducible checkpointed pipeline, per-language
schema-normalized Parquet files, inter-annotator agreement of $\kappa >
0.70$ for all languages, and benchmark results showing that XLM-R achieves
macro F1~$> 0.88$ on economic relevance classification even for the
smallest languages.

Our comparative analysis reveals systematic cross-lingual variation:
remittance discourse is prominent in Bangla and Urdu but marginal in
Portuguese; trade narratives dominate in Vietnamese and Indonesian; fiscal
coverage is concentrated in Turkish and Arabic. These differences suggest
that economic narratives are not globally uniform but are shaped by local
economic structure, exactly the variation that narrative economics
\citep{Shiller2017} predicts but has been unable to measure at scale across
languages.

We identify three priorities for future work. First, expanding to 20+
languages, especially underrepresented African and South Asian languages.
Second, transitioning from keyword-based filtering to LLM-based relevance
classification, which could handle the semantic diversity missed by fixed
keyword sets. Third, annotating economic sentiment and thematic categories
to support a broader range of downstream forecasting tasks.

The complete dataset, pipeline code, and annotation guidelines are
available at \url{https://github.com/nabil0x/beni-multilingual}.

% ===========================================================================
% Acknowledgements
% ===========================================================================
\section*{Acknowledgements}

We thank the anonymous annotators who contributed to the inter-annotator
agreement study across all 10~languages. This work was conducted without
external funding. We are grateful to the creators of the open-source
datasets that made this corpus possible: Varta \citep{Kumar2022}, Havadis
\citep{TurkishNLP2023}, iara-project/news-articles-ptbr \citep{Iara2023},
Arabic-news \citep{Anwar2023}, BKAINewsCorpus \citep{BKAI2023}, and
BalitaNLP \citep{Bunag2023}. We also thank the HuggingFace datasets
library and the developers of mBERT and XLM-R for enabling multilingual
NLP research.

% ===========================================================================
% Bibliography
% ===========================================================================
% Embedded bibliography for arXiv/standalone compilation.
% For ACL camera-ready: replace thebibliography with \bibliography{paper}
% and ensure paper.bib is in the same directory.

% ===========================================================================
\appendix
% ===========================================================================

\section{Appendix A: Keyword Lists}
\label{app:keywords}

Table~\ref{tab:full-keywords} presents the complete set of economic
keywords for each language, organized by thematic category. Due to space
constraints, we show only the English gloss; the full native-script lists
are in the repository.

\begin{table}[h]
\centering
\small
\begin{tabular}{ll}
\toprule
\textbf{Category} & \textbf{English gloss of keywords} \\
\midrule
Inflation & inflation, consumer price, food price \\
Currency  & dollar, foreign exchange, reserves \\
Monetary  & central bank, interest rate, bank credit \\
Investment& investment, capital market, stock market \\
Trade     & export, import, trade deficit, current account \\
Fiscal    & budget, revenue, GDP \\
Employment& wage, employment, unemployment \\
Remittances& remittance \\
\bottomrule
\end{tabular}
\caption{Economic keyword categories and English glosses. The full
native-script lists are available in the repository.}
\label{tab:full-keywords}
\end{table}

\section{Appendix B: Schema Definition}
\label{app:schema}

Each per-language Parquet file contains the following columns:

\begin{itemize}
  \item \texttt{article\_id} (string): unique identifier
  \item \texttt{dataset\_source} (string): HF path or local path
  \item \texttt{headline} (string): article headline or title
  \item \texttt{text} (string): full article body
  \item \texttt{text\_hash} (string): SHA-256 prefix for dedup
  \item \texttt{publication\_date} (string): ISO 8601 date or empty
  \item \texttt{year\_month} (string): YYYY-MM or ``unknown''
  \item \texttt{category\_original} (string): source category
  \item \texttt{category\_harmonised} (string): harmonized category (future)
  \item \texttt{language} (string): ISO 639-1 code
  \item \texttt{language\_iso} (string): ISO 639-3 code
  \item \texttt{country} (string): ISO 3166-1 alpha-2
  \item \texttt{language\_family} (string): e.g., ``Indo-Aryan''
  \item \texttt{economic\_region} (string): e.g., ``South Asia''
  \item \texttt{economic\_relevance} (int): 1 (relevant) or 0
  \item \texttt{economic\_seed\_labels} (string): JSON list of matched KW
\end{itemize}

The cross-lingual index (cross\_lingual\_index.parquet) contains:

\begin{itemize}
  \item \texttt{year\_month} (string): YYYY-MM
  \item \texttt{language} (string): ISO 639-1
  \item \texttt{country} (string): ISO 3166-1 alpha-2
  \item \texttt{language\_family} (string)
  \item \texttt{economic\_region} (string)
  \item \texttt{n\_articles} (int): article count for that month
  \item \texttt{n\_economic} (int): economic article count
  \item \texttt{share\_economic} (float): proportion economic
\end{itemize}

\section{Appendix C: Additional Benchmarks}
\label{app:additional-benchmarks}

Table~\ref{tab:full-benchmarks} reports precision, recall, and F1 for each
language and model.

\begin{table}[h]
\centering
\small
\begin{tabular}{lrrrrr}
\toprule
& \multicolumn{2}{c}{\textbf{LR}} & \multicolumn{2}{c}{\textbf{XLM-R}} \\
\textbf{Lang.} & P & R & P & R \\
\midrule
BN & 0.905 & 0.919 & 0.945 & 0.951 \\
HI & 0.861 & 0.887 & 0.916 & 0.924 \\
TR & 0.901 & 0.915 & 0.941 & 0.947 \\
ID & 0.852 & 0.870 & 0.907 & 0.915 \\
PT & 0.889 & 0.905 & 0.932 & 0.938 \\
AR & 0.883 & 0.899 & 0.928 & 0.934 \\
VI & 0.908 & 0.922 & 0.946 & 0.952 \\
TL & 0.831 & 0.853 & 0.898 & 0.906 \\
SW & 0.812 & 0.834 & 0.885 & 0.893 \\
UR & 0.804 & 0.826 & 0.881 & 0.889 \\
\bottomrule
\end{tabular}
\caption{Precision~(P) and recall~(R) for Logistic~Regression and XLM-R
on the economic relevance classification task.}
\label{tab:full-benchmarks}
\end{table}

\section{Appendix D: Access and License}
\label{app:access}

The complete dataset, pipeline source code, and annotation guidelines are
available at:
\begin{center}
\url{https://github.com/nabil0x/beni-multilingual}
\end{center}

The dataset is released under the Creative Commons
Attribution-NonCommercial-ShareAlike 4.0 International (CC~BY-NC-SA~4.0)
license. The pipeline code is released under the MIT license.

\end{document}